# The pH-dependent electrical potential and elastic modulus of a nanoscale DNA film and the resultant bending signal for a microcantilever biosensor

Neng-Hui Zhang[a, b,*], Wei-Lie Meng[b], Cheng-Yin Zhang[b], Jun-Zheng Wu[b]

a Department of Mechanics, College of Sciences, Shanghai University, Shanghai 200444, China
b Shanghai Key Laboratory of Mechanics in Energy Engineering, Shanghai Institute of Applied Mathematics and Mechanics, Shanghai University, Shanghai 200072, China

**Abstract**

The electrical/mechanical properties of nanoscale DNA films on solid substrates have a close relation with various detection signals of maicro-/nano-devices, such as bending deflection, frequency shift and indentation stiffness. However, the self-adjusting in microstructures and constituents due to surrounding fluctuations makes a DNA film exhibiting diverse properties, which brings a great difficulty in characterizing the relationship between the response signals and the detection conditions. This paper devotes to formulating several multiscale models to study the effect of pH-dependent ionic inhomogeneity on the electrical potential distributions and the graded elastic properties of a nanoscale DNA film and the related bending deflections of a microcantilever biosensor. First, the Langmuir isotherm was used to improve the classical Poisson-Boltzmann equation for polyelectrolyte solutions by introducing a new solution parameters to consider the effect of the inhomogeneous distribution of hydrogen ions on the electrical potential. Second, inspired by the Parsegian's mesoscopic attraction potential for cation condensed DNAs, the graded distribution properties of the particles were taken in the construction of an alternative interaction potential for both attraction-dominated and repulsion-dominated films. The new model parameters were obtained by curve fitting with the bending deflection experiments of a microcantilever done by Arroyo-Hernandez et al. Third, by the improved interaction potential and the thought experiment about the compression of a DNA bar in the context of macroscopic continuum mechanics, we investigated the diversity of elastic properties of single-stranded DNA (ssDNA) films due to the self-adaptability of its microstructures and constituents to the pH value, salt concentration and cation valence. Numerical results show that electronegative DNA could be overcharged with a positive electrical potential when the particle distribution conforms with the condition that the sign of the new introduced solution parameter is negative. There exists a transition from the pH-sensitive interval to pH-insensitive one for bending signals and elastic moduli. Negative elastic modulus is first revealed in the attraction-dominated ssDNA film constrained on the Au-layer of a microcantilever biosensor.

# 1. Introduction

The field of cantilever-based sensing is still relatively young and was initiated in the mid-1990s [1]. Due to its comparative advantages such as label-free, exquisite mass resolution, portable, cheap, highly parallel and fast response for field use [2], considerable interest has been attracted from applications such as point of care diagnostics, homeland security and environmental monitoring[3]. However, mirco-/nano-mechanical sensing has not yet been accepted as a practical alternative to well-established bioanalytical techniques such as ELISA, microarrays or electrophoresis methods [4].

DNA-related mirco-/nano-cantilever experiments show that the biosensing signals, i.e. deflections or frequencies, rely on changes in interactions among biomolecules and substrates, which could be incurred by many factors, such as fragment length, base pair sequence, concentration, packing density, hybridization density of DNA molecules, salt concentration and valence, pH value, refractive index in buffer solutions, surface charge of gold layer, surface preparation technology, the substrate properties, humidity, time, and temperature [5, 6].

As for the pH-related experiments, Shu et al. [7] found that the direction and amplitude of artificial DNA motor-induced cantilever motion was tuneable via control of buffer pH and ionic strength, and a sharp transition in the compressive surface stresses (i.e. downward bending) were observed at approximately pH 6.7. Zhang et al. [8] demonstrated that at low pH 4.5, hydration and electrostatic forces led to tensile surface stress (i.e. upward bending), implying the reduced accessibility of the bound single-stranded DNA (ssDNA) probe for hybridization, whereas at high pH 8.5, higher electrostatic repulsive forces bent the microcantilever downwards to provide more space for the target DNA, thus the hybridization efficiency peaks between pH 7.5 and 8.5. Watari et al. [9] studied the molecular basis of stress generation in aqueous environments focusing on the pH titration of mercaptohexadecanoic acid self-assembled monolayers on microcantilevers, and found that a tensile surface stress of +1.2 ± 0.3 mN/m at pH 6.0 was generated, conversely, the magnitude of compressive surface stress was found to increase progressively with pH ≥ 7.0, reaching a maximum of −14.5 ± 0.5 mN/m at pH 9.0, attributed to the enhanced electrostatic repulsion between deprotonated carboxylic acid groups. Johansson et al. [10] presented an SU-8 cantilever chip with integrated piezoresistors for detection of surface stress changes due to adsorption of biomolecules such as model mercaptohexanol, and the cantilevers were observed to bend up towards the Au-side for both increasing and decreasing pH values from 5.6 to 10.0. Similar behavior has been observed for Au-coated $SiO_2$ cantilevers [11]. Calleja et al. [12] and Johansson et al. [10] showed that SU-8 cantilevers could provide the enhanced stability to pH variations that could screen the molecular recognition signal.

However, there are few quantitative theories to interpret the above-mentioned pH-dependent upward/downward motions, which is one of key controversies in the community of mechanical sensors. Based on a coarse-grained DNA cylinder model, the Parsegian's mesoscopic repulsion potential [13] and its improved versions [6] revealing microscale interactions such as electrostatic force, hydration force and conformational entropy among DNA chains, water molecules, and salt ions, have been used successfully to predict the cantilever deflections [14, 15], frequencies [16], and the signal-related elastic properties of DNA films [5, 6, 17, 18] in the case of compressive surface stresses. But it might not be applied in the case of tensile surface stresses due to the negligence of the attractive interactions. Recently its improved version has been updated to interpret the attractive forces in the homogeneous cation condensed double-stranded DNA (dsDNA) solutions by combining single-molecule magnetic tweezers and osmotic stress experiments [19]. By introducing a three-component lattice model consisting of polymer, water, and vacancies, Wagman et al. [20] explained qualitatively the conversion mechanism between the tensile and compressive surface stresses when self-assembled monolayers of single-stranded DNA or PNA were exposed to water vapor that was found in Mertens et al.'s experiment [21]. However, the above-mentioned models did not consider the effect of pH value. By combining the modified Stoney's equation, Poisson-Boltzmann equation (PBE), and molecular dynamics (MD), Sushko et al. [22] developed a multiscale quantitative model to interpret the variations of the differential surface stress for nanomechanical biosensors with the variations in the pH value and temperature of buffer solutions, and the chain length of alkanethiol self-assembled monolayers (SAMs). But the computational cost of MD simulations for SAM thermal/mechanical parameters is high.

The paper is devoted to studying the effect of pH value on the electrical/mechanical properties of DNA films on substrates and the resultant deflection signals of DNA-microcantilevers. First, considering the inhomogeneous distribution of phosphate groups of DNA chains, H$^+$, OH$^-$, salt ions induced by the substrate effect, the pH-related hydrogel theory [23] and Langmuir isotherm are referred to present an alternative nonlinear PBE for a near-surface system to reveal the relation between the film potential and the surrounding solution conditions, and the analytical prediction of the electrical potential and the distribution rule of the net charge density for DNA solutions are given based on the linear PBE. Second, by the above acquired inhomogeneous properties of the net charge density and the attraction potential obtained by combining single-molecule magnetic tweezers and osmotic stress for multivalent cation condensed DNA assemblies [19], an alternative interaction potential is presented to reveal the effect of inhomogeneous charge density on the competition between microscopic attractive and repulsive interactions for DNA solutions on substrates. And the theorem of minimum energy and an thought experiment are used to formulate the analytical models for the bending deflections of DNA-microcantilevers and the elastic moduli of DNA films. Third, the electrical potential distributions of DNA films are discussed numerically according to three kinds of linear solutions. In the meantime, the empirical parameters in the present interaction potential are obtained by curve fitting with the upward and downward bending deflection experiments [24]. Then the effect of the pH value, and the salt valence and concentration on the deflections of microcantilevers and the moduli of DNA films are studied elaborately.

## 2. Mathematical models

*2.1 Electrical potential of DNA solution*

*2.1.1 Nonlinear PBE*

For highly charged polymers such as DNA, the classical nonlinear PBE, as a mean-field description, can have considerable limitations arising from the reduced structural detail with which the DNA macromolecule is approximated. To offset these problem, the inhomogenous distribution of particles across the thickness due to the substrate effect will be considered to update the classical nonlinear PBE. The DNA solution on a substrate contains fixed DNA chains, movable cations and anions, and water molecules, and it is a typical polyelectrolyte solution. Here, in the context of the mean field, water is viewed as a continuous medium with a high dielectric constant $\zeta$, and the solution electrical potential $\psi$ is determined by the charge density of the movable salt ions $\rho_m$ and that of the fixed DNA phosphate groups $\rho_f$, i.e., the electrical potential conforms with the following Poisson's equation [6]

$$\zeta \frac{d^2\psi}{dz^2} = -(\rho_f + \rho_m), \tag{1}$$

in which

$$\rho_m = q\sum_{i=1}^{n} z_i n_i, \tag{2}$$

where the coordinate axis $z$ is taken along the thickness of the DNA film, $i$ represents the $i$-th salt ion, H$^+$ or OH$^-$ ion, $q$ is the electric quantity of a proton, and $q = 1.6 \times 10^{-19}$ C, $z_i$ the valence of each ion, $n_i$ the charge number density of each ion, which satisfies the Boltzmann distribution

$$n_i = n_i^{\infty} \exp(-z_i q \beta \psi), \tag{3}$$

in which $n_i^{\infty}$ is the bulk concentration of each ion, $\beta = 1/k_B T$, where $k_B$ is the Boltzmann constant, and $k_B = 1.38 \times 10^{-23}$ J/K, $T$ is the temperature. Note as two kinds of free ions, H$^+$ and OH$^-$ also follow the Boltzmann distribution [29], i.e.

$$n_{H^+} = n_{H^+}^{\infty} \exp(q\beta\psi), \tag{4a}$$

$$n_{OH^-} = n_{OH^-}^{\infty} \exp(-q\beta\psi), \tag{4b}$$

where $n_{H^+}^{\infty}$ and $n_{OH^-}^{\infty}$ are the bulk densities of H$^+$ and OH$^-$, respectively. At room temperature, their corresponding molar concentrations satisfy $c_{H^+}^{\infty} c_{OH^-}^{\infty} = 1 \times 10^{-14}$ M$^2$.

In Eq. (1), the charge density of the phosphate groups fixed on DNA chains is given as

$$\rho_f = z_f n_f q, \tag{5}$$

in which $z_f$ is the valence of the fixed groups, $n_f$ is the charge number density of the fixed groups, which accords with the following Langmuir isotherm [25]

$$n_f = n_0^s \alpha, \tag{6a}$$

$$\alpha = \frac{K}{V_r(K + c_{H^+})}, \qquad (6b)$$

where $\alpha$ reflects the ionization degree of the fixed groups, $K$ is the dissociation constant of the fixed groups, and $K = 0.01\,\text{M}$ for the phosphate groups [26]; $n_0^s$ is the total concentration of the ionizable groups in the DNA film at the initial state, and $n_0^s = \eta N / h_p$, here $\eta$ is the packing density of DNA chains, $N$ is the nucleotide number, $h_p$ is the thickness of the DNA film. Obviously $n_0^s$ is closely related to the packing conditions. The molar concentration of $H^+$ is given as $c_{H^+} = n_{H^+}/(1000 N_A)$, which determines the pH value of the solution, i.e. $\text{pH} = -\log_{10} c_{H^+}$, here $N_A$ is the Avogadro constant. The local hydration constant of the DNA film $V_r$ is defined as the ratio of fluid volume $V_f$ to solid volume $V_s$, i.e. $V_r = V_f / V_s$. As shown in Fig.1, in the case of uniform hexagonal packing condition, the local hydration constant is given as

$$V_r = (3\sqrt{3} r_D^2 - 2\pi r_d^2)/(2\pi r_d^2) \qquad (7)$$

where $r_D$ is the side length of the hexagonal cell, and $r_D = \sqrt{2/(\eta\sqrt{3})}/2$; $r_d$ is the radius of the DNA cylinder, corresponding to 1 nm for dsDNA and 0.5 nm for ssDNA. Here the coarse-grained cylinder model for DNA chains is taken as in the previous works [6, 13].

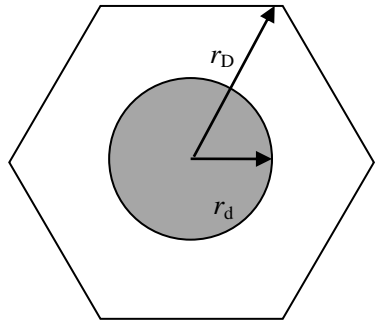

Fig. 1 Schematic of the model cell in the case of hexagonal packing condition

For convenience, the DNA solution on a substrate is divided into two areas: area I contains water, ions and DNA chains, and area II only contains water and ions. From Eqs. (1)−(5), we can derive an improved nonlinear PBE as

$$\zeta \frac{d^2\psi_I}{dz^2} = -\{\sum_{i=1}^{n} z_i n_i^{\infty} q \exp(-z_i q \beta \psi_I) + \frac{z_f q n_0^s K}{V_r[K + n_{H^+}^{\infty} \exp(-z_f q \beta \psi_I)]}\}, 0 < z \leq h_p, \quad (8a)$$

$$\zeta \frac{d^2\psi_{II}}{dz^2} = -\sum_{i=1}^{n} z_i n_i^{\infty} q \exp(-z_i q \beta \psi_{II}), \quad z > h_p. \quad (8b)$$

Note that here the Boltzmann assumption is taken for the ion distributions in order to capture the near-surface properties of the DNA film on a substrate, which is a little different from the numerical simulation based on the Nernst-Planck equation for pH-stimulus-responsive hydrogels [23, 25]. In addition, from Eqs. (4a), (6a), (6b), (7), we know that the dissociation degree of the fixed groups is a function of the distance $z$, which can reflect the related findings by the dissociation gradient model obtained from the full free-energy minimization and the experimental observations for the polyacrylic acid brush [27].

*2.1.2 Linear PBE and its solution*

When the term $-z_i q \beta \psi \ll 1$, the exponential terms in Eqs. (8a) and (8b) could be expanded in a Taylor series, retaining only the first two terms [28]. Due to the bulk electroneutrality, i.e. $\sum_{i=1}^{n} z_i n_i^{\infty} q = 0$, a linear PBE is derived as

$$\frac{d^2\psi_I}{dz^2} = \text{sgn}(f) \kappa_1^2 \psi_I - z_f q n_0^s \alpha_0 / \zeta, \quad 0 < z \leq h_p, \quad (9a)$$

$$\frac{d^2\psi_{II}}{dz^2} = \kappa_0^2 \psi_{II}, \quad z > h_p, \quad (9b)$$

where $f = \sum_{i=1}^{n} n_i^{\infty} z_i^2 - \frac{z_f^2 n_0^s c_{H^+}^{\infty} K}{V_r(K + c_{H^+}^{\infty})^2}$, and $\alpha_0 = \frac{K}{V_r(K + c_{H^+}^{\infty})}$ is a parameter related to the ionization degree of the fixed phosphate group in the bulk solution [29], $\kappa_0 = (q^2 \beta \sum n_i^{\infty} z_i^2 / \zeta)^{1/2}$ is the classical inverse Debye screening length, $\kappa_1 = (q^2 \beta |f| / \zeta)^{1/2}$ is a new introduced parameter, which reveals the effect of inhomogeneous properties of the fixed charges on the solution parameters for a near-surface system. Note that the solution parameters $\kappa_1$ and $\kappa_0$ for areas I and II, respectively, are different. It seems to be similar to Ohshima and Makino's formula [30], however, they are totally different in essential. Ohshima and Makino [30] adopted the homogenous assumption of the fixed charges, in which the difference of solution parameters between the two areas came from the different concentrations of salt ions, so it is in terms of the classical inverse Debye screening length. Whereas we adopt an inhomogeneous assumption of the fixed charges, which brings about the changes of the buffer solution parameter not only in the amounts but also in the sign.

If the dimensionless parameters and quantities are introduced as follows

$$\Psi_{\mathrm{I}} = \frac{q\psi_{\mathrm{I}}}{k_{\mathrm{B}}T}, \quad \Psi_{\mathrm{II}} = \frac{q\psi_{\mathrm{II}}}{k_{\mathrm{B}}T}, \quad Z = \kappa_1 z, \quad H = \kappa_1 h_{\mathrm{p}}, \quad \bar{n}_{\mathrm{f}} = \frac{z_{\mathrm{f}} q^2 n_0^s \alpha_0}{\zeta \kappa_1^2 k_{\mathrm{B}} T}, \quad \gamma = \kappa_0 / \kappa_1,$$

then the linear PBE (9) could be transformed into

$$\frac{\mathrm{d}^2 \Psi_{\mathrm{I}}}{\mathrm{d}Z^2} = \mathrm{sgn}(f)\Psi_{\mathrm{I}} - \bar{n}_{\mathrm{f}}, \quad 0 \leq Z \leq H, \tag{10a}$$

$$\frac{\mathrm{d}^2 \Psi_{\mathrm{II}}}{\mathrm{d}Z^2} = \gamma^2 \Psi_{\mathrm{II}}, \quad Z > H. \tag{10b}$$

The boundary and continuous conditions are given as [30]

$$\left.\frac{\mathrm{d}\Psi_{\mathrm{I}}(Z)}{\mathrm{d}Z}\right|_{Z=0} = 0, \quad \Psi_{\mathrm{II}}(Z)\big|_{Z \to \infty} = 0, \tag{11}$$

$$\Psi_{\mathrm{I}}(Z)\big|_{Z=H} = \Psi_{\mathrm{II}}(Z)\big|_{Z=H}, \quad \left.\frac{\mathrm{d}\Psi_{\mathrm{I}}(Z)}{\mathrm{d}Z}\right|_{Z=H} = \left.\frac{\mathrm{d}\Psi_{\mathrm{II}}(Z)}{\mathrm{d}Z}\right|_{Z=H}. \tag{12}$$

From the above linear PBE (10a) and (10b), the boundary conditions (11) and the continuous conditions (12), the electrical potential distributions could be obtained for areas I and II, respectively. Next, according to the sign of the parameter *f*, the discussion is divided into the following three parts:

**Case I** If $f > 0$, then

$$\Psi_{\mathrm{I}} = 2C_1 \cosh(Z) + \bar{n}_{\mathrm{f}}, \quad C_1 = \frac{\exp(H)\bar{n}_{\mathrm{f}} \gamma}{1 - \gamma - \exp(2H)(1+\gamma)}, \tag{13a}$$

$$\Psi_{\mathrm{II}} = C_2 \exp(-\gamma Z), \quad C_2 = \frac{\exp(\gamma H)[\exp(2H) - 1]\bar{n}_{\mathrm{f}}}{-1 + \gamma + \exp(2H)(1+\gamma)}. \tag{13b}$$

If neglecting the effect of $H^+$, or taking the homogenous assumption of $H^+$ concentration, i.e. $\gamma = 1$, Eq. (13a) degenerates into

$$\Psi_I = \bar{n}_f[1 - \exp(-H)\cosh(Z)]. \tag{14}$$

Obviously, Eq. (14) shows a hyperbolic cosine property for the potential distribution, which has the same expression predicted by Ohshima and Makino's model [30]. From Eq. (13a), we know

$$Z = 0, \quad \Psi_I = \bar{n}_f \left[ \frac{2\exp(H)\gamma}{1 - \gamma - \exp(2H)(1+\gamma)} + 1 \right] < 0;$$

$$Z = H, \quad \Psi_I = \frac{\exp(2H) - 1}{-1 + \gamma + \exp(2H)(1+\gamma)} \bar{n}_f < 0,$$

$$\Delta\Psi = \Psi(H) - \Psi(0) = \bar{n}_f \frac{\gamma[2\exp(-H) - \exp(-2H)]}{\gamma - 1 + (\gamma + 1)\exp(2H)} > 0.$$

Hence the potential is always negative in the DNA film, and the top potential is bigger than that the bottom one, which agrees with most of the mean-field theories [31] and numerical results [27, 32, 33].

**Case II** If $f = 0$, then

$$\Psi_I = -\bar{n}_f' Z'^2 / 2 + C_1, \quad C_1 = \bar{n}_f' H'(H'/2 + 1), \tag{15a}$$

$$\Psi_{II} = C_2 \exp(-Z'), \quad C_2 = \bar{n}_f' H' \exp H'. \tag{15b}$$

where due to $\kappa_1 = 0$, we have to introduce some new dimensionless parameters and quantities as follows: $Z' = \kappa_0 z$, $H' = \kappa_0 h_p$, $\bar{n}_f' = \frac{z_f q^2 n_0^s \alpha_0}{\zeta \kappa_0^2 k_B T}$. From Eq. (15a), we know

$$Z' = 0, \Psi_I = \bar{n}_f' H'(H'/2 + 1) < 0; \quad Z' = H', \Psi_I = \bar{n}_f' H' < 0,$$

$$\Delta\Psi = \Psi(H') - \Psi(0) = -\bar{n}_f' H'^2 / 2 > 0.$$

Similar to **Case I**, the potential is always negative in the DNA film. As shown in Eq. (15a), the second type of the solution is a typical quadric form, which is consistent with the form derived by Zhuilina and Borisov [29] by the variational approach.

**Case III** If $f < 0$, then

$$\Psi_I = C_1 \cos(Z) - \bar{n}_f, \quad C_1 = \frac{\bar{n}_f \gamma}{\gamma \cos H - \sin H}, \tag{16a}$$

$$\Psi_{\mathrm{II}} = C_2 \exp(-\gamma Z), \quad C_2 = \frac{\sin H}{\gamma \cos H - \sin H} \exp(\gamma H) \bar{n}_{\mathrm{f}}. \tag{16b}$$

From Eq. (16a), different from **Case I** and **Case II**, here $\Psi_{\mathrm{I}}(0)$ and $\Psi_{\mathrm{I}}(H)$ might be positive or negative depending on the buffer solution conditions. Hence, in **Case III**, the potential difference between the top and the bottom of the DNA film is likely to be positive. The relevant electrical experiments and the MD or DFT simulations have shown that the electrical potential of the DNA solution has the overcharged phenomenon on the special solution conditions. However, the previous predictions of the electrical potential based on the classical PBE are always negative, which could not be relied on explaining the overcharged phenomenon [34]. Here the graded distribution properties of the fixed phosphate group, movable hydrogen and hydroxyl ions are included, which makes the positive film potential possible at some special conditions.

In a brief summary, by considering the inhomogeneous distribution of the fixed group, the present improved PBE could be used to reveal the inhomogeneous effect of hydrogen ion on the solution parameters. This kind of graded properties will change not only the quantity but also the sign of the characteristic parameters, which further alters the solution structure. When $f > 0$, the film potential demonstrates with a hyperbolic cosine distribution, the model could degenerate the previous model under the hypothesis of homogenous surface charge. When $f = 0$, it demonstrates with a parabolic distribution. When $f < 0$, it demonstrates with a cosine distribution, in which the positive potential is possible. In addition, it can be seen from Eqs. (5)−(7), different from Zhulina and Borisov [29], the local hydration constant [35] defined by Li et al. [25] is introduced here to make the prediction of overcharged phenomenon possible at special solution conditions.

*2.2 Interaction potential of DNA film*

Based on the above analysis of the electrical potential, we will discuss the distributions of particles and effective charges, and further improve the Parsegian's attraction potential for multivalent cation condensed DNA assemblies [19] with the inclusion of the graded charge distributions.

*2.2.1 Distribution functions of particles and effective charges*

The equation of state for a thermodynamic system is given as [36]

$$\Pi = \frac{nRT}{V} = cRT, \tag{17}$$

in which $\Pi$ is the osmotic pressure for a dilute solution, $V$ the solution volume, $c$ the solution concentration, $R$ the ideal gas constant, $n$ the amount of substance for the solute, $T$ the absolute temperature. It can be seen from Eq. (17), at given temperature, the osmotic pressure is proportional to the particle number, but not the type and the size of the solute. For a near surface system, Eq. (17) should be adapted due to the prominent effect of inhomogeneity.

Substituting Eqs. (2)−(6) into the right side of Eq. (1) yields the effective charge density

$$\rho_e = \rho_m + \rho_f = q[\sum_{i=1}^{n} z_i n_i^{\infty} \exp(-z_i q \beta \psi) + n_0^s \alpha], \tag{18}$$

and the total particle density

$$n_t = \sum_{i=1}^{n} n_i + n_f = \sum_{i=1}^{n} n_i^{\infty} \exp(-z_i q \beta \psi) + n_0^s \alpha. \tag{19}$$

Obviously, for a multivalent salt, $\rho_e \neq n_t q$, i.e., the distribution function of the effective charge along the film thickness does not coincide with that of the solute including salts and DNA chains, whereas for a monovalent salt, they have the same rule.

Furthermore, on the condition of the weak potential, by expanding Eqs. (18) and (19) into the Taylor series, and noticing the bulk electroneutrality, we could obtain the effective charge density and the total particle density, respectively,

$$\rho_e = fq^2 \beta \psi + b_0, \tag{20}$$

$$n_t = a_1 q \beta \psi + b_0, \tag{21}$$

in which

$$a_1 = \frac{z_f n_0^s n_{H^+}^{\infty} K}{V_r (K + n_{H^+}^{\infty})^2}, \quad b_0 = \frac{z_f q n_0^s K}{V_r (K + n_{H^+}^{\infty})}.$$

It can be seen from the comparison of Eqs. (20) and (21), in the case of weak potential, the effective charge density and the particle density have the similar distribution rules across the film thickness.

*2.2.2 Osmotic pressure and free energy of DNA film*

By the experiments of single-molecule magnetic tweezers and osmotic stress, Todd et al. [19] proposed that the osmotic pressure in DNA bulk solutions based on the cylinder model could be decomposed into the following two parts, i.e.

$$\Pi = -Ae^{-d/\lambda_D} + Be^{-2d/\lambda_D}, \tag{22}$$

in which $d$ is the interchain distance, $\lambda_D$ is the characteristic decay length for intermolecular interactions between DNA molecules. For attraction-dominated cases, and $\lambda_D = 0.46$ nm [19], whereas for repulsion-dominated cases, $\lambda_D = 1/\kappa_0$, in which the ionic effect should be considered [13]. And the first term in the right side of Eq. (22) is induced by the attraction interactions, the second term is induced by the repulsion interactions, and *A* and *B* are the undetermined empirical parameters. At given temperature, Eq. (17) shows that the osmotic pressure is proportional to the particle number, so with the consideration of the inhomogeneous effect, the osmotic pressure of the DNA film on a substrate can be assumed as

$$\Pi = \bar{f}(z)(-Ae^{-d/\lambda_D} + Be^{-2d/\lambda_D}), \tag{23}$$

where $\bar{f}(z)$ is the distribution function of the particle number density across the film thickness. From Eqs. (13a), (15a), (16a) and (21), and by the definition of the dimensionless parameters and quantities, the distribution functions of the particle number density are obtained as if $f > 0$, $\bar{f}(z) = \cosh(\kappa_1 z)$; if $f = 0$, $\bar{f}(z) = 1$; if $f < 0$, $\bar{f}(z) = \cos(\kappa_1 z)$

Once the expression of the osmotic pressure is obtained, the free energy density in the DNA film could be derived as [19]

$$\begin{aligned} W_b &= -\int_\infty^d \Pi \sqrt{3} d\mathrm{d}d \\ &= \sqrt{3}\bar{f}(z)[-A(\lambda^2 d + \lambda)e^{-d/\lambda} + B(\frac{\lambda^2}{4} + \frac{\lambda}{2}d)e^{-2d/\lambda}]. \end{aligned} \tag{24}$$

Note that the competition between attraction and repulsion parts is included in the formula (24), which is similar to the Purohit's expression [37] considering only the repulsion part. Different from Todd et al. [19], the inhomogeneous effect in the near surface system is included in Eq. (24). Different from the previous methodology in constructing the repulsion potential for DNA systems [6, 13], the geometrical hypothesis of the infinite cylinder is abandoned, and the relation (24) is more fit for short DNA systems used in microcantilever-based detections. In the case of the hexagonal packing pattern [38], the total mesoscopic free energy of the DNA film is

$$W_B = \int_0^{h_p} 3lb\eta W_b dz,  \qquad (25)$$

in which $l$ and $b$ are the length and width of the film, respectively, $h_p$ is the thickness of the film, which could be predicted by the following improved formula [6, 29]

$$h_p = KL_c \eta^\mu I^{-\nu},  \qquad (26)$$

in which the contour length $L_c = Na$, $N$ the nucleotide number, $a$ the nucleotide length, corresponding to $0.34 \pm 0.04$ nm for dsDNA and $0.22 \pm 0.04$ nm for ssDNA [40], $K = 0.00089$, $\mu$ and $\nu$ are the undetermined constant, $I$ the total ion strength [8], and $I = \frac{1}{2}\sum_{i=1}^n c_i^\infty z_i^2$, where $c_i^\infty$ is the molar concentration of the ion $i$, and the sum is calculated over all ions in the solution.

*2.3 Bending deflection of DNA-microcantilever*

In the process of biodetections based on the static-mode of microcantilevers, the adsorption of DNA chains on the substrate forms a soft matter film stored with an amount of free energy that will make the deformable substrate bending upwards or downwards. The change of bending deflections could be monitored by the optical or electrical methods. The identification of the relationship between the sensitive signals and the environmental factors calls the demand of quantitative models [41]. Among all kinds of macroscopic/microscopic quantitative models, the multiscale models based on the Parsegian's mesoscopic interaction potential are preferred due their low cost of computation and high predictability of microscale details. Here, the above improved mesoscopic free energy (i.e. Eq. (24) or (25)) will be used to formulate a multiscale analytical model for the bending deflection of a DAN-microcantilever.

The cantilever substrate usually is a laminated structure consisting of Au-, Ti-/Cr- and Si-layers. During the biodetection, the stored mechanical energy of the substrate in the context of macroscopic continuum mechanics could be described by Zhang's two-variable method, i.e. [6, 42]

$$W_M = A_1 \varepsilon_0^2 + B_1 \varepsilon_0 \kappa + C_1 \kappa^2, \tag{27}$$

in which $\varepsilon_0$ is the normal strain at the interface between the film and the substrate, $\kappa$ is the curvature of the neutral axis, and

$$A_1 = bl(E_{Au}h_{Au} - E_{Cr}h_{Cr} + E_{Cr}h_{Cr} + E_{Si}h_{Si})/2,$$

$$B_1 = bl[-E_{Au}h_{Au}^2 - E_{Cr}h_{Cr}(2h_{Au} + h_{Cr}) - E_{Si}h_{Si}(2h_{Au} + 2h_{Cr} + h_{Si})]/2,$$

$$C_1 = bl\{E_{Au}h_{Au}^3 + E_{Cr}[(h_{Au} + h_{Cr})^3 - h_{Au}^3] + E_{Si}[(h_{Au} + h_{Cr} + h_{Si})^3 - (h_{Au} + h_{Cr})^3]\}/6.$$

where $E$ and $h$ are the elastic modulus and the thickness of each layer, and $b$ and $l$ are the width and the length of the substrate, respectively.

The continuous condition at the interface requires the interchain distance after the bending to satisfy [15]

$$d(z) = (1 + \varepsilon_0 + \kappa z)d_0 \tag{28}$$

where $d_0$ is the initial interchain distance depending on the packing density, and for the hexagonal packing pattern, $\eta = 2/(\sqrt{3}d_0^2)$.

To obtain an analytical form of the mesoscopic free energy stored in the DNA film in the case of small formation, with the help of Eq. (28), the free energy density in Eq. (24) could be expanded into the Taylor series in terms of small quantities, $\varepsilon_0$ and $t = \kappa z$, i.e.

$$W_b(\varepsilon_0, t) \approx A_2' + B_2'\varepsilon_0 + C_2't + D_2'\varepsilon_0^2 + E_2'\varepsilon_0 t + F_2't^2, \tag{29}$$

in which

$$A_2' = W_b \bigg|_{\substack{\varepsilon_0=0 \\ t=0}}, \quad B_2' = \frac{\partial W_b}{\partial \varepsilon_0}\bigg|_{\substack{\varepsilon_0=0 \\ t=0}}, \quad C_2' = \frac{\partial W_b}{\partial t}\bigg|_{\substack{\varepsilon_0=0 \\ t=0}},$$

$$D_2' = \frac{\partial^2 W_b}{\partial \varepsilon_0^2}\bigg|_{\substack{\varepsilon_0=0 \\ t=0}}, \quad E_2' = \frac{\partial W_b}{\partial \varepsilon_0 \partial t}\bigg|_{\substack{\varepsilon_0=0 \\ t=0}}, \quad F_2' = \frac{\partial^2 W_b}{\partial t^2}\bigg|_{\substack{\varepsilon_0=0 \\ t=0}}.$$

Substituting Eq. (29) into Eq. (25) yields the total free energy in the DNA film

$$W_B \approx A_2 + B_2\varepsilon_0 + C_2\kappa + D_2\varepsilon_0^2 + E_2\varepsilon_0\kappa + F_2\kappa^2, \tag{30}$$

in which

$$A_2 = 3lb\eta A'_2 h_p, \quad B_2 = 3lb\eta B'_2 h_p, \quad C_2 = 3lb\eta C'_2 h_p^2 / 2,$$

$$D_2 = 3lb\eta D'_2 h_p, \quad E_2 = 3lb\eta E'_2 h_p^2 / 2, \quad F_2 = lb\eta F'_2 h_p^3.$$

By using the principle of minimum energy, $\delta W_T = \delta W_B + \delta W_M = 0$, the microcantilever deflection are given as

$$w = -\kappa x^2 / 2, \tag{31}$$

in which

$$\kappa = \frac{2C_2 A_1 - B_1 B_2 + 2C_2 D_2 - B_2 E_2}{B_1^2 - 4A_1 C_1 - 4D_2 C_1 + 2E_2 B_1 + E_2^2 - 4F_2 A_1 - 4D_2 F_2}.$$

Note that the constants $A, B, \mu, \nu$ in Eq. (30) will be identified by the relevant experiments. It can be seen from Eqs. (24), (26) and (28) that the surrounding fluctuations will induce the changes not only in the local particle distribution and the local deformation of the film, but also in the global deformation of the cantilever. In another word, here we present a multiscale mathematical model to identify the coupling relation between the local microstructure variation and the global deformation of the cantilever. The adaptability of the local microstructure for this kind of soft matter is not elaborated distinctly in the previous attraction potential model [19].

*2.4 Elastic modulus of DNA film*

Mechanical properties of DNA films on substrates have great influences on biodetection signals [6]. In this section, the DNA film will be viewed as a bar in the context of macroscopic continuum mechanics. Based on the thought experiment about a biaxial iso-strain compression, if knowing the free energy density of the DNA film as shown in Eq. (24), the principle of energy conservation is used to obtained the following effective normal stress

$$\sigma_x = \frac{3}{2} \frac{d[W_b \eta|_{d=d_0(1-\varepsilon_0)}]}{d\varepsilon_0}, \tag{32}$$

In the case of small deformation, the elastic modulus depends on the initial slope of the stress-strain curve of the DNA film, i.e.

$$E = \frac{\partial \sigma_x}{\partial \varepsilon_0}\bigg|_{\varepsilon_0 = 0}. \tag{33}$$

# 3. Results and discussion

*3.1 Distributions of electrical potential and net charge density*

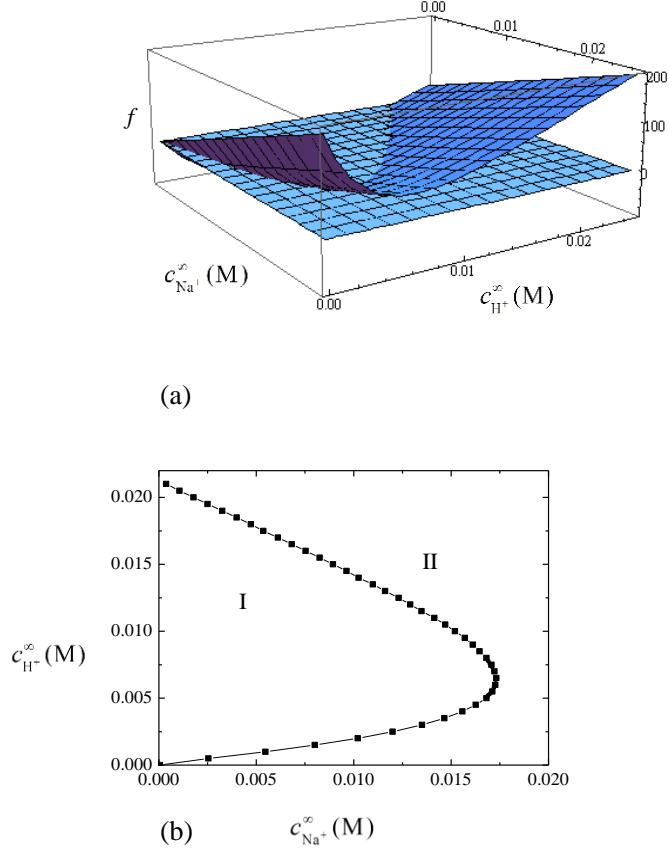

(a)

(b)

Fig. 2 The relation between the solution parameter and the concentrations of hydrogen and sodium ions (a) 3D plot; (b) zero equipotential curve

As mentioned in the above discussions, the salt solution parameter $f$ is a predominant factor for the electrical potential distribution that will give us valuable information in order to adapt the mesoscopic free energy with the inclusion of the inhomogeneous distributions of solute particles. Take NaCl solution, for example, Figs. 2a and 2b shows the effect of $Na^+$ and $H^+$ concentrations. Note the concentrations of other ions are given as: $c_{OH^-}^\infty = 10^{-14}/c_{H^+}^\infty$, $c_{Cl^-}^\infty = c_{Na^+}^\infty + c_{H^+}^\infty - c_{OH^-}^\infty$. As shown in Fig. 2b, the parameter plane is divided into two areas. When the parameters locate in area II, $f > 0$; whereas in area I, $f < 0$ due to the predominance of phosphate groups on DNA chains; under some special combinations of $c_{Na^+}^\infty$ and $c_{H^+}^\infty$, $f = 0$, namely, zero equipotential curve as shown Fig. 2b. For these three cases, as shown in Eqs.

(13a), (15a) and (16a), different distribution functions should be chosen in the construction of an inhomogeneous free energy in the DNA film. In the following section, different distributions will be examined at different solution conditions. The parameters related to the dsDNA solution are taken as $T = 298\,\text{K}$, $\zeta = 81 \times 8.854 \times 10^{-12}\,\text{F/m}$, $N = 20\,\text{nt}$, $\mu = 0.17$, $\nu = 0.18$, $\eta = 0.09\,\text{chain/nm}^2$ [6].

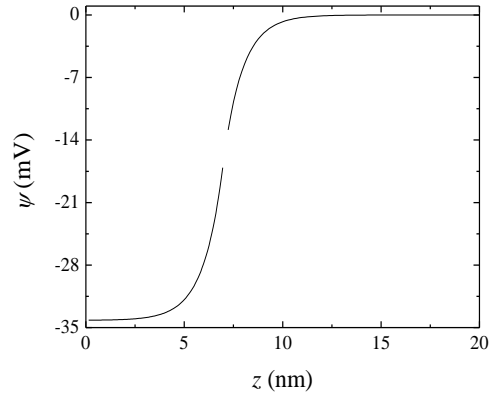

(a)

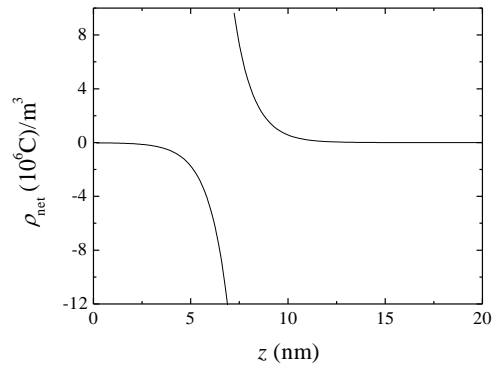

(b)

Fig. 3 The distributions across the thickness for **Case I**, (a) the electrical potential; (b) the net charge density

(i) If $c_{\text{Na}^+}^\infty = c_{\text{Cl}^-}^\infty = 0.1\,\text{M}$, $c_{\text{H}^+}^\infty = c_{\text{OH}^-}^\infty = 10^{-7}\,\text{M}$, then, Eqs. (13a) and (13b) are used to predict the electrical potential distribution as shown in Fig. 3a. Obviously, wherever inside or outside the DNA film, the electrical potential is always negative. Inside the film, the bottom potential is lower than the top one. This is caused by the hydrolyzation of phosphate groups on DNA chains, which brings about negative charges attracting cations and repelling anions. This inhomogeneous interactions make the interface electrical potential transiting to zero not suddenly, but smoothly, which is consistent with the predictions by other mean-field models [29, 31] and various numerical simulations [27, 32, 33].

Next, the net charge density will be studied and the result is shown in Fig. 3b. Inside the film, the net charge density varies monotonically from zero to the minimum, whereas outside the film, it decreases from the maximum to zero. Hence, this typical distribution will trigger a dipole at the interface between the film and the free solution, which is accordance with DFT simulations [27]. The dipole is caused by the inhomogeneous adsorption of cations with DNA fixed groups. Inside the film, the neutralization of numerous cations with DNA negative charges makes the film bottom ($z < 5$ nm) almost in an electroneutrality state. However, at the interface, the ability of adsorption weakens, the cation number gets smaller, and it is not enough to neutralize DNA negative charges, and this results in an electronegativity state in the film top area. Outside the film, numerous cations collect in the free solution close the interface due to the attraction of DNA negative charges, this results in an electropositivity state, and with the increase of the distance to the interface, the attraction weakens and the collection decreases, this results in an approach to a state of zero potential.

(ii) On condition 1 (i.e. $c_{\text{Na}^+}^\infty = 0.01\,\text{M}$, $c_{\text{Cl}^-}^\infty \approx 0.012\,\text{M}$, $c_{\text{H}^+}^\infty \approx 0.0019\,\text{M}$, $c_{\text{OH}^{-1}}^\infty \approx 5.2 \times 10^{-12}\,\text{M}$) or condition 2 (i.e. $c_{\text{Na}^+}^\infty = 0.01\,\text{M}$, $c_{\text{Cl}^-}^\infty \approx 0.024\,\text{M}$, $c_{\text{H}^+}^\infty \approx 0.014\,\text{M}$, $c_{\text{OH}^{-1}}^\infty \approx 7 \times 10^{-13}\,\text{M}$), $\text{sgn}(f) = 0$, Eqs. (15a) and (15b) are used to predict the electrical potential distributions as shown in Fig. 4a. Similar to **Case I**, the electrical potential is always negative everywhere, and the absolute potential attains its maximum at the film bottom. Different from **Case I**, first, with the decrease of the salt concentration, the electrostatic screening effect of salt ions on DNA charged groups weakens, so the DNA film gets thicker; second, the electrical potential demonstrates a typical a parabolic distribution in **Case II**, whereas a hyperbolic cosine one in **Case I**.

As for the distribution of the net charge density as shown in Fig. 4b for **Case II**, it is totally different from **Case I**. The net charge density keeps a constant inside the film, which could be easily deduced from Poisson's equation with a parabolic distribution of the electrical potential as shown in Eq. (15a). It is a special case, where the concentrations of hydrogen ions and salt ions satisfy a special condition, i.e. $f = 0$. Inside the DNA film, the competition between the phosphate groups on DNA chains, salt ions and hydrogen ions make the net charge density to be a constant; whereas outside the film, the cations collect at the interface due to the attraction of DNA negative charges, and this makes the net charge density positive in the free solution.

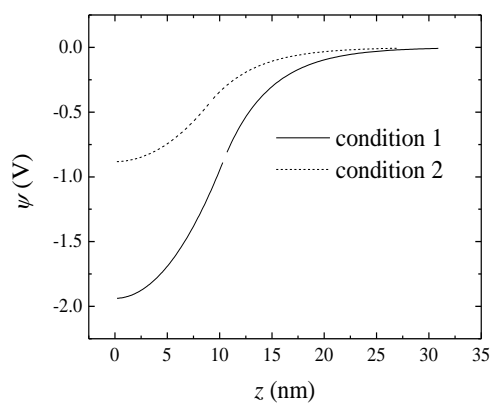

(a)

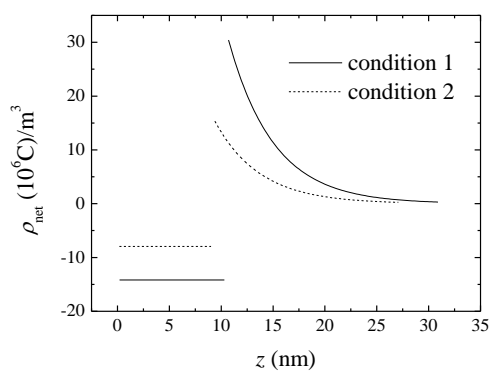

(b)

Fig. 4 The distributions across the thickness for **Case II**, (a) the electrical potential; (b) the net charge density

(iii) If $c_{Na^+}^\infty = 10^{-3}$ M, $c_{Cl^-}^\infty = 0.011$ M, $c_{H^+}^\infty = 10^{-2}$ M, $c_{OH^-}^\infty = 10^{-12}$ M, then $\text{sgn}(f) < 0$, Eqs. (16a) and (16b) are used to predict the electrical potential distribution as shown in Fig. 5a. Different from **Cases I** and **II**, the electrical potential in **Case III** is positive in the film bottom area, whereas negative in the film top area closing to the interface, and outside of the film, it approaches to zero with the increase of the distance to the interface. It is the first time to predict a positive potential in the bottom of DNA film by our improved PBE (i.e. Eqs. (9a) and

(9b)) with the inclusion of the inhomogeneous effect across the thickness. This overcharged phenomenon [34] is a result of the competition between a large amount of hydrogen ions and a small amount of salt ions. The success of predicting this special potential distribution might give us a chance to characterize attractive interactions based on the coarse-grained DNA cylinder models.

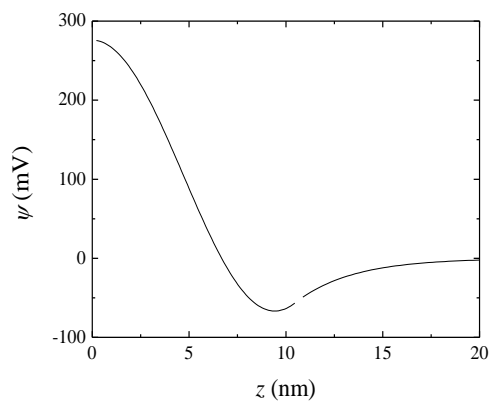

(a)

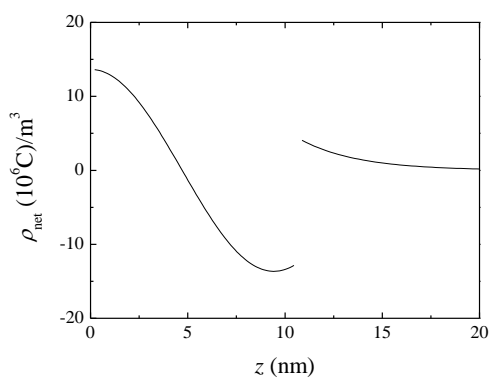

(b)

Fig. 5  The distributions across the thickness for **Case III**, (a) the electrical potential; (b) the net charge density

In this case, as shown in Fig. 5b, the net charge density shows an interesting distribution pattern. Inside the DNA film, the net charge density is positive in the bottom area, whereas negative in the top area. The positive net charge in the film is caused by the collection of hydrogen ions and cations in the bottom area and the escape of anions from the film. It is interesting for the positive net charge emerging at the local area of the film, because the originally negative phosphate groups on DNA chains are neutralized by salt cations and hydrogen ions. Hence, this special salt solution condition requires amounts of hydrogen ions, which provides us a possibility of capturing the inverse sign of the net charge by a mean-filed theory.

*3.2 Bending deflections of ssDNA-microcantilevers*

*3.2.1 Curve fitting with immobilization experiments*

If the property of the particle distribution is known, then what is its effect on the microcantilever deflection? In the immobilization experiments done by Arroyo-Hernandez at al. [24], the microcantilever deflection bends upwards or downwards, which was supposed to be related to the charged state of the surface Au-layer prepared by the resistive (Res) or e-beam (Eb) evaporation techniques. We could not resort to the Parsegian's repulsion potential [13] or its updated versions [6, 43] to elucidate the mechanism of upward deflections because the attractive interaction was omitted in the previous models. Here as the solution parameter $f > 0$, the hyperbolic cosine distribution function as shown in Eq. (13) and the prediction model for microcantilever deflection as shown in Eq. (31) are used to obtain the empirical parameters $A, B, \mu, \nu$. The substrate parameters are given as $l = 400\,\mu\text{m}$, $b = 100\,\mu\text{m}$, $h_{\text{Si}} = 1\,\mu\text{m}$, $h_{\text{Cr}} = 2\,\text{nm}$, $h_{\text{Au}} = 20\,\text{nm}$, $E_{\text{Si}} = 180\,\text{GPa}$, $E_{\text{Cr}} = 279\,\text{GPa}$, $E_{\text{Au}} = 73\,\text{GPa}$. The DNA parameters are taken as $N = 16\,\text{nt}$, $\eta = 0.35$ chain/nm$^2$ at the Res condition, or $\eta = 0.417$ chain/nm$^2$ at the Eb condition [44]. In a PBS buffer solution (137 mM NaCl, 2.7 mM KCl, 8 mM Na$_2$HPO$_4$, 2 mM KH$_2$PO$_4$; pH = 7.5), $T = 298\,\text{K}$, $c_{\text{Na}^+}^\infty = 0.153\,\text{M}$, $c_{\text{Cl}^-}^\infty = 0.1397\,\text{M}$, $c_{\text{H}^+}^\infty = 10^{-7.5}\,\text{M}$, $c_{\text{OH}^-}^\infty = 10^{-6.5}\,\text{M}$; the effect of other ions is omitted due to their tiny quantities.

As shown in Fig. 6, at the Res packing condition, the tensile surface stress (i.e. upward bending) corresponds to the attraction-dominated ssDNA film, whereas at the Eb packing condition, the compressive surface stress (i.e. downward bending) corresponds to the repulsion-dominated ssDNA film. Obviously, the fitting curves agree well with the experiments. The empirical parameters for ssDNA films are obtained as $A = 3200\,\text{MPa}$, $B = 0.25\,\text{MPa}$, $\mu = 0.16$, $\nu = 0.23$ with a determination coefficient of 0.999 at the Res condition, and

$A = 0.25\,\text{MPa}$, $B = 6686\,\text{MPa}$, $\mu = 0.17$, $\nu = 0.24$ with a determination coefficient of 0.997 at the Eb condition, respectively. With the above fitting parameters and Eq. (26), the ssDNA film thicknesses are predicted as 4.9 nm at the Eb condition, which approaches to the AFM measured value of 4.2 nm in the discharged state by Arroyo-Hernández et al. [24]. And the predicted thickness at the Res condition is 3.2 nm, which approaches to the AFM measured value of 3.4 nm in air or in liquid by Legay et al. [35]. In addition, the contracting tendency of the DNA film at the Res condition is in accordance with the observations in DNA condensation experiments [19].

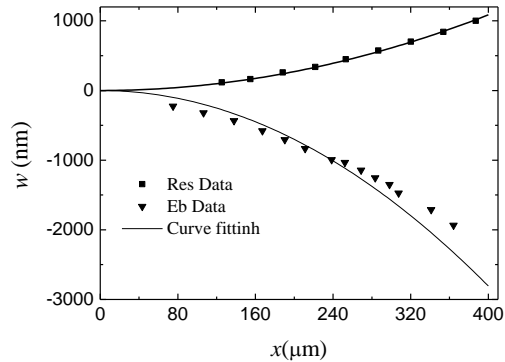

Fig. 6 Curve fitting with immobilization deflections of ssDNA-microcantilever

*3.2.2 Prediction of immobilization deflections*

In Fig. 7, the computational parameters are taken as $\eta = 0.25\,\text{chain/nm}^2$, and $c_{\text{cation}} = 1\,\text{mM}$ for the attraction-dominated case (i.e. at the Res condition), or $c_{\text{cation}} = 0.1\,\text{M}$ for the repulsion-dominated case (i.e. at the Eb condition), the other parameters are the same as those in Fig. 6.

First, we will study the effect of the pH value. As for the repulsion-dominated case, small downward deflections occur when pH < 2, but a sharp transition emerges when $\text{pH} \in (2, 4)$, and reaching a maximum bending signal at pH 4. This is resulted from the accumulation of $\text{OH}^-$ within the ssDNA film with the increase of the pH value that strengthens the repulsion interaction between the negative phosphate groups on DNA chains and $\text{OH}^-$ ions in the buffer solution. However, the saturation of $\text{OH}^-$ will make the repulsion reaching its maximum. The variation tendency with the pH value is in agreement with the experimental observation done by

Shu at al. [7], who observed a downward bending from pH 5 to pH 9 and a maximum bending signal for DNA motors at approximately pH 6.7. Whereas in the attraction-dominated case, the similar variation tendency of upward deflections could be observed with the increase of the pH value. In fact, during the preparation of Au-layer by the Res evaporation technique, the surface charge is more positive [24], which makes a big difference in the empirical parameters *A* and *B*. So with the increase of the pH value or OH$^-$, the attraction between the positive Au-layer and OH$^-$ ions will be strengthened, which causes more bigger upward deflections. However, the predicted tendency here is opposite to the dsDNA experimental results obtained by Zhang et al. [8], which might arise from different packing conditions. As we know, dsDNA doubles negative charges compared with the corresponding ssDNA. So the increase of OH$^-$ in the dsDNA film will strengthen the repulsion part more efficiently rather than the attraction one and it makes the upward deflections smaller. Another common characteristic for both kinds of ssDNA films should be noted that the deflections at acid conditions are always smaller than that at alkaline conditions, which agrees well the experiments done by Shu et al. [7] and Zhang et al. [8].

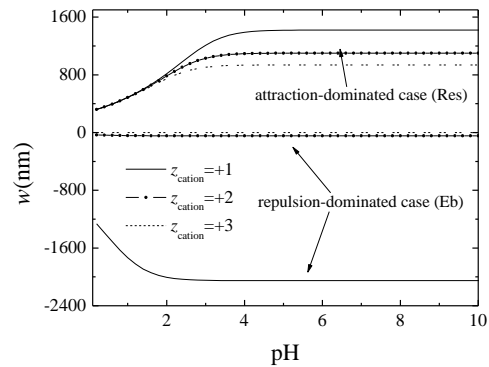

Fig. 7 Effect of pH value and cation valence on immobilization deflections of ssDNA-microcantilever

Next, we will study the effect of cation valence. As for the repulsion-dominated case, the increase of cation valence will decrease the downward deflections because it is more efficient for cations with higher valences to neutralize the negative phosphate groups on DNA chains, and this results in a prominent decrease of the repulsion part, which has been validated by the osmotic experiment on cation condensed dsDNA done by Todd et al. [19]. Whereas in the attraction-dominated case, the increase of cation valence will diminish the upward deflections, because the increase of cation valence will neutralize more negative phosphate groups and further weaken the attraction between the positive Au-layer and the negative charges. However, an enhancing attraction with the increase of the cation valence has been observed in the single-molecule magnetic tweezers experiment of dsDNA [8]. The difference also arises from different surrounding conditions. In the attraction-dominated ssDNA film, the positive charges are provided from not only the cations in the buffer solution but also the Au-layer, and the attraction comes from the interaction among DNA molecules, salt ions and Au-layer, so the competition of microscopic interactions is more complicated than that in the single-molecule magnetic tweezers experiment. In addition, the repulsion-dominated film is more sensitive to the variation of the valence from 1 to 2 than the attraction-dominated film does, however the trend turns into a reversal when the valence varies from 2 to 3. The difference comes from different solution parameters. Note the characteristic decay length is cation-independent for the attraction-dominated DNA [19], whereas cation-dependent for the repulsion-dominated DNA [46].

*3.3 Elastic modulus of DNA films*

*3.3.1 Effect of pH* value *and cation concentration*

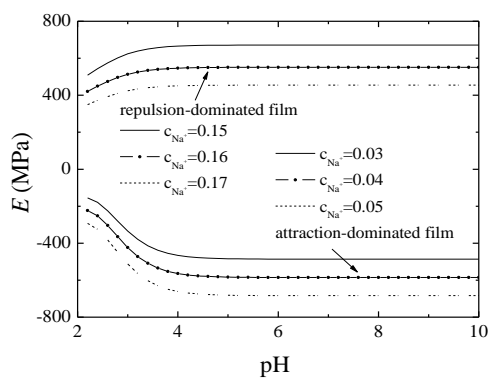

Fig. 8 Effect of pH value and cation concentration on elastic modulus of ssDNA film

From the above analyses, we know that the microcantilever bends upwards or downwards depending on the specific packing conditions. What it does mean to the film mechanical properties? Because the mechanical properties of the adsorbate film are closely related to the signals in static-/dynamic-mode detections based on microcantilevers. Especially in the dynamic-mode detection, the elastic modulus of an adsorbate film is one of key parameters to be identified, because the stiffness effect cannot be neglected and results in complex frequency signals. In Fig. 8, the computational parameters is taken as $\eta = 0.17 \text{ chain/nm}^2$ for the attraction-dominated film, or $\eta = 0.1 \text{ chain/nm}^2$ for the repulsion-dominated film, and $z_1 = +1$, the other parameters are the same as those in Fig. 6.

As for the effect of the pH value, similar to the variation tendency of deflections for both repulsion-dominated and attraction-dominated cases, with the increase of the pH value, the elastic modulus is enhanced from a small value, and attains its maximum at about pH 4.0, then keeps a constant after pH 4.0, which is also caused by strengthening the attraction interaction between the $OH^-$ ions and the positive Au-layer at the Res condition or the repulsion interaction between the $OH^-$ ions and negative phosphate groups at the Eb condition. Note here the order of the predicted elastic modulus is no more than $10^3$ MPa, which falls into the scope of experimental observations done by Legay et al. [45]. In their AFM experiments using the Sneddon model, the elastic modulus of ssDNA in air is from 0.1 to $10^4$ MPa. In addition, with the increase of cation concentration, the attraction-dominated film increases its modulus at lower concentration, whereas the repulsion-dominated film decreases at higher concentration. A similar phenomenon has been found in a microcantilever experiment done by Wu et al. [47], in which the change in hybridization deflection increased sharply before 0.1 M while decreased smoothly after 0.1 M. Our previous Monte Carlo simulation based on the Parsegian's repulsion potential [13] also validated this point that there is a critical salt concentration for the elastic modulus of ssDNA film, by which the effect of salt ions is divided into the two domains: the salt-sensitive area and the salt-insensitive area.

It is worthy to point out that the attraction-dominated film has a negative elastic modulus. Although the negative stiffness effect has been found by the experiments for some special composite materials [48, 49] and is also proved by the related theoretical analyses [50, 51], it is the first time for us to reveal theoretically the negative modulus for the attraction-dominated ssDNA film. According to Lakes' analysis, negative stiffness entails a reversal of the usual codirectional relationship between force and displacement in deformed objects, which may result in superior mechanical/electrical properties, such as extreme damping, stiffness and large wave vector. Here the ssDNA immobilized on the positive-stiffness substrate has stored the chemical energy, which stabilizes the kind of composite materials containing DNA, water and salt ions, because negative stiffness structures and materials are unstable by themselves.

*3.3.2 Effect of cation valence and size effect*

Next, we will study the cation valence effect on the elastic modulus distribution across the ssDNA thickness as shown in Fig. 9. In computation, the parameters are taken as $c_{\text{cation}} = 30\text{ mM}$ for the attraction-dominated film, or $c_{\text{cation}} = 0.15\text{ M}$ for the repulsion-dominated film, and $\eta = 0.17\text{ chain/nm}^2$, $\text{pH} = 7$, the other parameters are the same as those in Fig. 6.

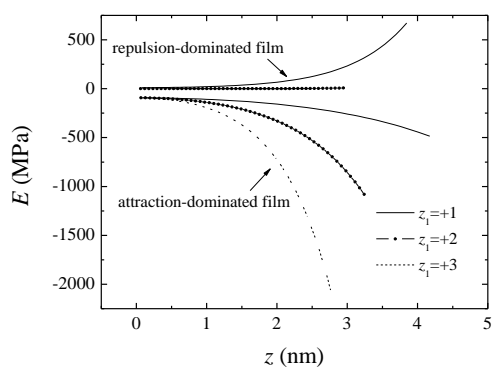

Fig. 9 Effect of cation valence on elastic modulus distribution of ssDNA film

First, for whatever nature of the film, the elastic modulus is not constant, but presents obviously an graded distribution over the full film thickness, which has also been observed in the AFM nanoindentation experiments in air by Legay et al. [45]. The gradient in the top region is rather sharp, taking $z_1 = +1$ for instance, the local modulus several times greater than the bottom value, with the absolute maximums of 490 MPa and 670 MPa for the repulsion-dominated and attraction-dominated films, respectively. Besides, similar properties could be found in other kinds of soft materials, such as biological organics, like virus, cell, nanoparticles, whose shells are stiffer outside and softer inside. This gradient mechanical property is important for biomaterials because the outer shell is supposed to be tough enough to resist more attacks from environment while the interior should be as soft as possible for physiological activities [52].

Second, with the increase of the cation valence, the film thickness shrinks, and the absolute modulus goes up for the attraction-dominated film, while drops down for the repulsion-dominated film. Especially for the repulsion-dominated film when $z_1 = +2 \text{ or } +3$, the modulus becomes zero and loses the ability to resist any deformation. This is also caused by the higher neutralization ability of the cation with a higher valence, which decreases the attraction at the Res condition or the repulsion at the Eb condition and the resultant deflection signals as shown in Fig. 7. However, it does not mean the modulus decrease. For example, it can be seen from Fig. 9 that the absolute modulus for the attraction-dominated film rises up reversely. In fact, from the definition as shown in Eq. (29), (32), (33), the modulus is related to the energy density, which depends on not only the whole free energy in the film but also the film volume. The increasing modulus is caused by the fact the reducing amount of the volume as a denominator overpasses that of the total free energy as a numerator. Obviously, the modulus-thickness relations at different packing conditions accord with not only the classical Halle-Petch effect (i.e., material strength increases with the decrease of size) [52] at the Eb condition, but also the Inverse Halle-Petch effect (i.e., material strength decreases with the decrease of size) at the Res condition.

## 4. Conclusions

Based on the classical Poisson-Boltzmann equation for polyelectrolyte solutions, an improved version is developed to characterize the relation between the pH-dependent ionic inhomogeneity and the electrical potential in a DNA solution on a solid substrate. The study shows that the competition among salt ions, $H^+$ and $OH^-$ will decide the sign of the new introduced solution parameter, namely, the distribution property of the particles, which makes it possible for the electronegative DNA to be overcharged with a positive electrical. It is the first time for an improved mean-field model to capture this interesting phenomenon that has been observed in the previous electrochemical experiment and the related MD and DFT simulations [34].

By the Parsegian's mesoscopic attraction potential for cation condensed DNAs, the change in the microscopic graded distribution of the particles across the DNA film is included in the construction of an alternative interaction potential for attraction-dominated or repulsion-dominated DNA films and the succeeding prediction of bending deflections of microcantilevers. Our theoretical curve fittings agree well with the deflection experiments done by Arroyo-Hernández et al. [24, 44]. Predictions show that, with the increase of the pH value, the enhanced attraction between $OH^-$ ions and the positive Au-layer in the attraction-dominated case or the enhanced repulsion between the negative phosphate groups and $OH^-$ ions in the repulsion-dominated case strengthens the bending signals for microcantilevers with whichever ssDNA films. These results qualitatively reproduced the apparent dependence of downward deflections on pH value seen in the microcantilever experiments done by Shu at al. [7]. There exists a transition between the pH-sensitive area and the pH-insensitive one, which offers us a chance to manipulate both the direction and amplitude of an array of autonomous microcantilever sensors in micromechanical machinery [7].

By the thought experiment about the compression of a ssDNA bar in the context of macroscopic continuum mechanics, a graded elastic property over the full thickness due to the inhomogeneous distribution of the particles is revealed. The film bottom is more softer than the top area where the modulus is about $0.1 \sim 10^3$ MPa. The predicted tendency across the film thickness and the order of the modulus agree well with the AFM experiments done by Legay et al. [45]. In addition, the flexibility of DNA solution microstructures to surrounding factors makes the DNA film exhibiting a remarkable size effect, which leads to a diversity in elastic modulus. On the acidic condition, the film is much softer with pH 4.0 as its critical value after which the modulus keeps steady. With the increase of cation concentration, the modulus increases for the attraction-dominated film at lower concentration, while decreases for the repulsion-dominated film at higher concentration. As the cation valence varies from 1 into 2, the repulsion-dominated film changes its stiffness into zero and loses the ability to resist any deformation. Furthermore, a big difference is that the attraction-dominated film has a negative modulus, which is the first time revealed in DNA systems. Such diverse mechanical properties in the variable environments are potentially useful and beneficial for further understanding of biomimetic film in interface tissue engineering [53, 54], sustainable design of nanoscale biosensors [28] and gene therapy for pathogenic viruses [55, 56].

## AUTHOR INFORMATION

**Corresponding Author**

*E-mail: nhzhang@shu.edu.cn.

**Notes**

The authors declare no competing financial interest.

## ACKNOWLEDGMENTS

The Natural Science Foundation of China (Nos. 11272193 and 10872121) and the Shanghai Pujiang Program (No. 15PJD016) are acknowledged for financial support.

**TOC Graphic**

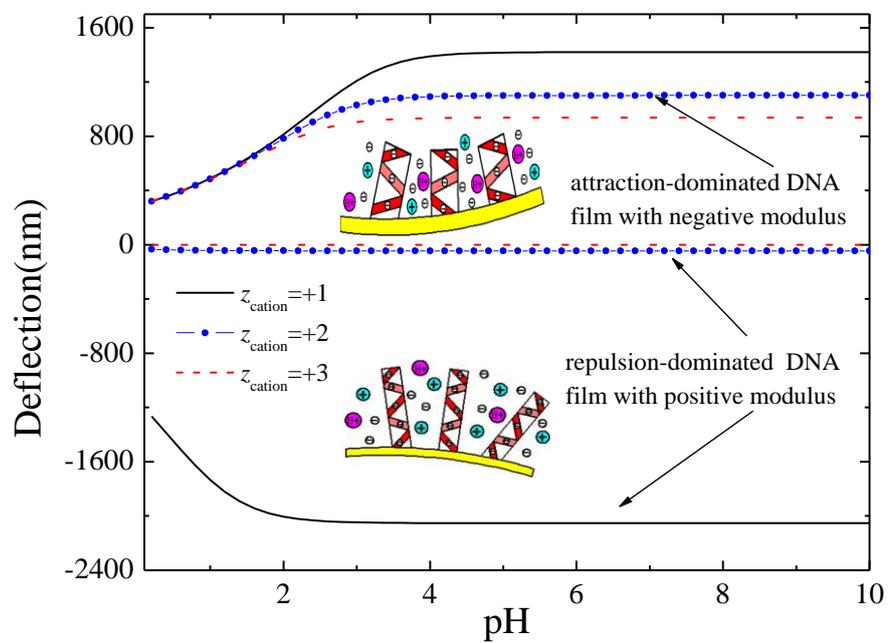